\documentclass[11pt,twoside]{article}

\usepackage{asp2006}
\usepackage{epsf}
\usepackage{psfig}
\usepackage{lscape}

\begin{document}

\title{The Stellar Observations Network Group -- the Prototype}   %%% Fill in title
%\author{F. Grundahl,\affil{Aarhus Universitet, Department of Physics and 
%Astronomy, Ny Munkegade, 8000 Aarhus C, Denmark } }
%\author{U. G. J{\o}rgensen\affil{NBI} }

\author{F. Grundahl$^{1}$, 
        J. Christensen-Dalsgaard$^{1}$, 
        H. Kjeldsen$^{1}$, 
        U.G. J{\o}rgensen$^{2}$,
        T. Arentoft$^{1}$,
        S. Frandsen$^{1}$, and
        P. Kj{\ae}rgaard$^{2}$
       }
\affil{$^{1}$Department of Physics and Astronomy,
Aarhus University, DK-8000 Aarhus C, Denmark }
\affil{$^{2}$Niels Bohr Institute, Copenhagen University,
DK-2100 Copenhagen, Denmark}

\begin{abstract} %%% Abstract to run on from here.
The Stellar Observations Network Group (SONG) has obtained full funding 
for the design, construction and implementation of a prototype telescope 
and instrumentation package
for the first network node. We describe the layout of such a node and its
instrumentation and expected performance for radial-velocity measurements. 
The instrumentation consists of
a 1m telescope, equipped with two cameras for photometry of microlensing 
events with the lucky-imaging technique and a high-resolution spectrograph
equipped with an iodine cell for obtaining high-precision radial velocities 
of solar-like stars, in order to do asteroseismology. The telescope will 
be located in a dome of $\sim4.5$m diameter, with two lucky-imaging cameras 
at one of the Nasmyth foci and the spectrograph and 
instrument control computers at a Coud{\'e} focus, located in an adjacent
container. 
Currently the prototype telescope and instrumentation are undergoing detailed
design. Installation at the first site (Tenerife) is expected during 
mid--late 2010, followed by extensive testing during 2011. 
\end{abstract}

%\keyword{Stellar Observations Network Group}
%\keyword{asteroseismology}
%\keyword{gravitational microlensing}
%\keyword{exoplanets}
%\keyword{instrumentation}

\section{Introduction}

 Stellar Observations Network Group (SONG) is an initiative started in 
Denmark to create a global network of highly specialized 
1m telescopes aimed at doing time-domain astronomy. In particular the 
goals are to: produce exquisite data for asteroseismic studies of stars
across most of the HR diagram (with focus on solar-like stars), and to 
search for, and characterize, the population of low-mass planets in orbit
around other stars via the microlensing and radial--velocity methods. 

 A ground-based network with a sufficient number of nodes will ensure  
observations with a high duty-cycle, needed to obtain stellar oscillation 
spectra with high frequency precision and without aliasing problems. 
Furthermore, the light-curve anomalies in microlensing events 
occur at unpredictable times and thus, to find and characterize these, 
near-continuous observations are needed.

 The group behind SONG has obtained funding for the design and 
construction of a full prototype network node which shall be completed 
in late 2011. In the following we describe some of the aspects of the
ongoing work and the expected performance of the prototype. 

%%%%%%%%%%%%%%%%%%%%%%%%%%%%%%%%%%%%%%%%%%%%%%%%%%%%%%%%%%%%%%%%%%%%%%%%%%%
\section{Observational goals}

Below we shall briefly account for the main goals and requirements that the
SONG instruments must fulfil in order to meet the requirements to 
do asteroseismology and measure microlensing light curves.

\subsection{Asteroseismology}

To study efficiently the oscillations of solar-like
stars, the best strategy for ground-based  observations is to measure 
the change in radial velocity of their surface. This requires a high radial 
velocity precision in the few  m/s range. 
For the best ground-based instruments 
this level of precision is now routine, and some are at the sub-meter per 
second precision level \citep{harps, iodine1}  
over short timescales. For the SONG project the aim is 
to achieve a velocity precision better than 1m/s for a $V=0$ star per 
minute of observation. This requires an efficient, high-resolution 
spectrograph and a CCD camera with a fast readout in order to allow a high
observing duty cycle. 

Our aim is to obtain a network duty cycle near 80\%. This will be achieved
by having 8 nodes distributed at existing northern and southern hemisphere 
sites, and well distributed in longitude. Experience from BiSON 
\citep{bison1}
show that it is realistic to reach this level with 6--8 stations, 
consistent with the results of \citet{mosser}.
It is clear that targets in the equatorial zone will have the highest degree
of coverage since these can be observed from both northern and southern 
nodes. With both northern and southern sites in the network, full-sky 
coverage will be possible. 

\subsection{Gravitational microlensing}

For the study of microlensing events towards the bulge of the Milky Way, 
SONG will employ the lucky-imaging method \citep{lucky1}. By observing the 
target field with a CCD camera which can read out at high speeds 
($\approx30$Hz), and then only selecting 
the images of best quality for subsequent co-addition, it is possible
to obtain images with high spatial resolution. Since the bulge fields are
in general quite crowded, this offers a big advantage in the achievable 
photometric precision and depth. Observations for this purpose only 
require a small field of view, and the current design foresees a field
of 46\arcsec$\times$46\arcsec.
The microlensing observations have similar requirements as the asteroseismic
observations with respect to the observing duty-cycle. \citet{lensing1} 
discusses the prospects for microlensing studies with SONG in more detail.

\subsection{Other possibilities}

One of the possibilities we are considering for SONG is to use the 
spectrograph to observe the oscillations of the Sun during daytime. This
is done by pointing the telescope to the blue sky and measuring the 
velocities in exactly the same way (through the iodine cell) as the 
stars are observed at night. In this way we will observe
the sun-as-a-star and complement the existing facilities (ground and 
space) for solar observations. See \citet{sun1} for an example of this
with the HARPS and UCLES spectrographs.

%%%%%%%%%%%%%%%%%%%%%%%%%%%%%%%%%%%%%%%%%%%%%%%%%%%%%%%%%%%%%%%%%%%%%%%%%%%
\section{Layout of the instrumentation}
%%%%%%%%%%%%%%%%%%%%%%%%%%%%%%%%%%%%%%%%%%%%%%%%%%%%%%%%%%%%%%%%%%%%%%%%%%%

The instrumentation at each site consists of a 1m telescope equipped with
a high-resolution spectrograph at a Coud{\'e} focus and two lucky-imaging
cameras at one of the two Nasmyth foci. This allows simultaneous two-colour 
imaging. 

\subsection{Dome and enclosure}
\begin{figure}[!ht]
\plotone{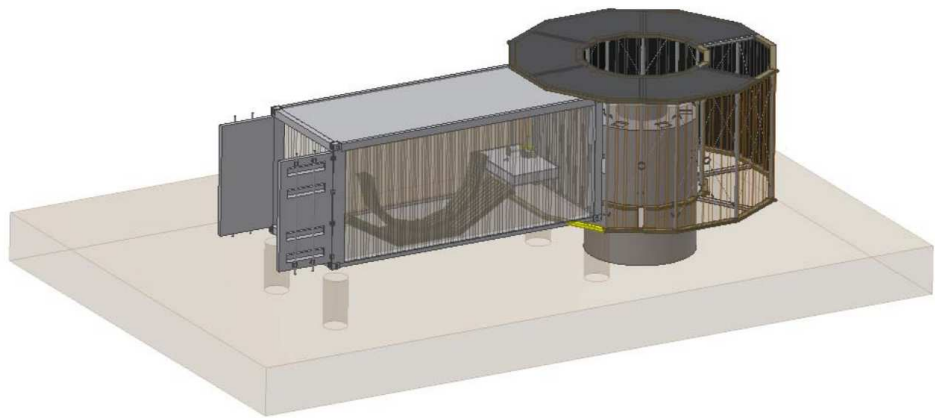}
\caption{
\label{fig:pier}
The basic structures of a SONG node, comprising a concrete pier
for supporting the telescope, a 20 foot container and a dome support 
structure. In the final version the structure carrying the dome will be 
approximately 1m higher with the dome floor at the level of the container
roof. Side-ports will be installed in the extra 1m height to improve 
nighttime ventilation of the dome. The inside of the container will be 
split in two compartments, one for the instruments nearest to the telescope
pier, and one at the other end for control computers and other electronics. }
\end{figure}

The telescope will be housed in a dome of approximately 4.5m diameter,
and the Coud{\'e} room will be located in a 20 foot shipping container with
a significant level of insulation. Computers and hard drives for observatory
control and data reduction will be located in this container as well. The 
use of a container for housing instrumentation and computers allows
cost savings and provides for a rather small footprint,
since the only permanent buildings needed will be the telescope concrete
pier and the footings on which the container is attached to the ground. A
separate concrete foundation will carry the support for the spectrograph such 
that it is mechanically de-coupled from the container. 
Figure~\ref{fig:pier} shows the general layout. 

Since any potential SONG node should be at an existing observatory, access 
to electrical power and internet will already be available. A weather station 
with a cloud monitor is also part of the instrumentation. 
The expected date for arrival of the telescope to the first site is mid-2010. 

\subsection{Telescope optics and imaging}

Lucky-imaging for the microlensing science will be carried out with two 
cameras located at one of the two Nasmyth stations. We decided to not 
place these at the Coud{\'e} focus since this would lead to less than optimal
image quality (many optical surfaces), lower overall efficiency and higher
costs. At the Nasmyth station a mirror slide with several positions will 
send the light to the two imaging cameras, or allow the light to pass on
to the Coud{\'e} train.  

For lucky-imaging it is important to have correction for atmospheric 
dispersion (ADC), as well as field de-rotation (the telescope is alt-az
mounted). The field de-rotation will be done with an optical de-rotator
of the Abbe-K{\"o}nig type for both imaging and spectroscopy. The imaging
at Nasmyth will be in two wavelength regions, with a split at 6700{\AA}. In
the ``short'' wavelength channel a beam splitter will be placed, which 
sends a small fraction of the light to a (cheap) CCD camera for continuous 
focus monitoring. In this way the telescope will always be at its optimal 
focus. Both of the two imaging channels will be equipped with filter wheels. 

The pixel scale for the lucky-imaging cameras is $\sim0\farcs09$ in order
to provide adequate sampling (in the red channel) for  nearly 
diffraction-limited imaging. At a seeing of 1\arcsec\, FWHM, the 1m SONG
telescope has $D/r_0$ of 5--6  at a wavelength of 8000{\AA}. In this regime 
the lucky-imaging method works extremely well, which implies that a very 
significant improvement in image quality can be expected via the use of
this method. At the best observing sites, imaging with a FWHM close to 
0\farcs3 can be expected. 

We should note that the CCD cameras will of course also offer the 
possibility to do photometry of other objects, such as variable stars 
and gamma-ray bursts. 

\subsection{The Coud{\'e} train and spectrograph focal plane}

\begin{figure}[!ht]
\plotone{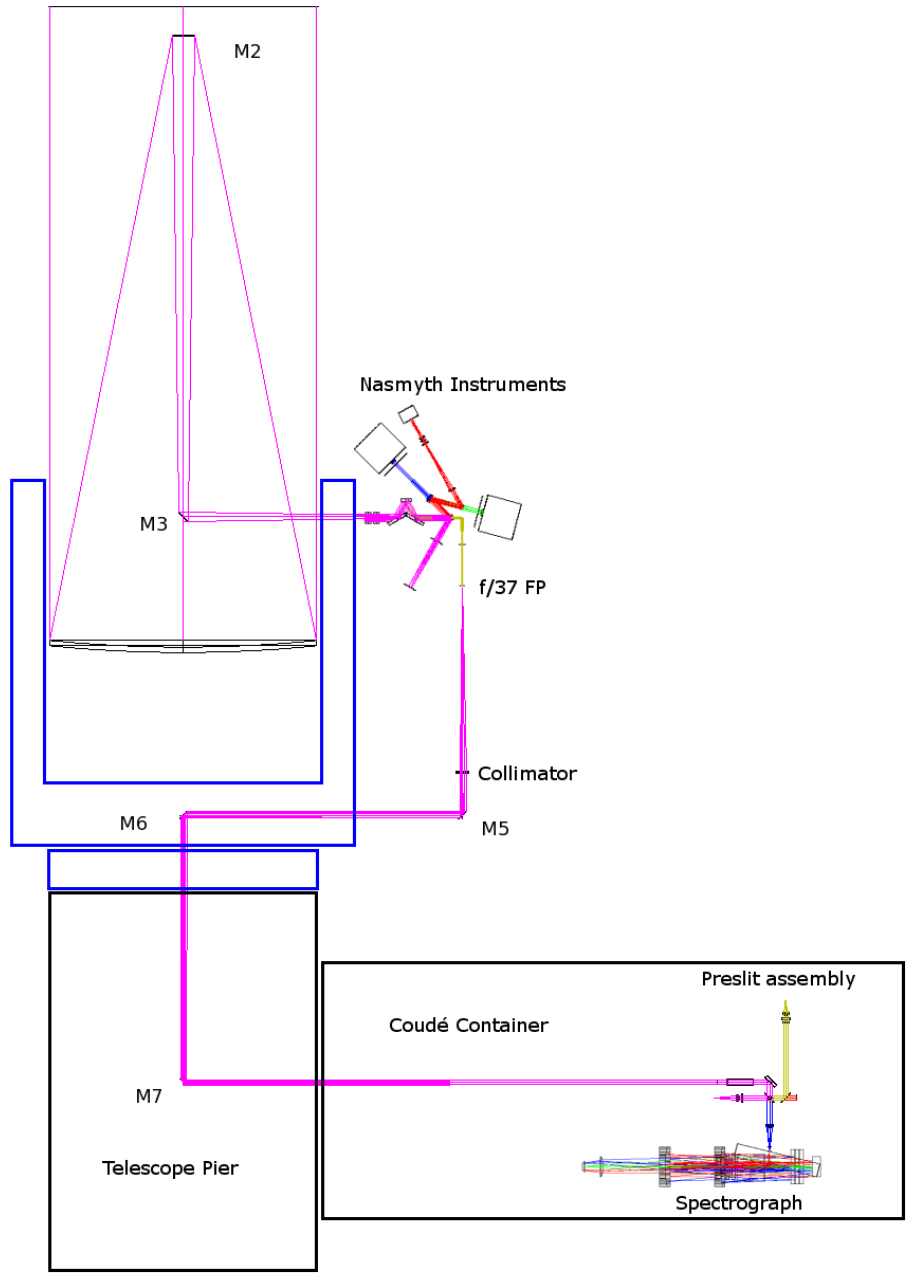}
\caption{
\label{fig:optical_layout}
A schematic layout for the SONG prototype node. The relative sizes
of the components are not completely correct, but only intended to show the
layout. }
\end{figure}

The second main instrument for SONG is the spectrograph which is located
at a Coud{\'e} focus. By removing the mirror that directs the light to the
Nasmyth station the light from M3 will instead go to other mirrors 
(M4, M5, M6, M7 and M8, see Fig.~\ref{fig:optical_layout}) and ultimately 
end in the instrument
container where the spectrograph is located. Radial velocities are measured
using the iodine technique \citep{iodine1}. This implies that only a limited
wavelength range, 5000--6200{\AA} is needed (discussed in more 
detail below). With such a short wavelength interval to be ``transported''
to the Coud{\'e} focus, highly optimized anti-reflective and 
reflective coatings can be used. This results in the total efficiency of
the Coud{\'e} train being nearly 95\%. 

At the pre-slit assembly several functions will be available; these include
calibration light (ThAr and halogen lamps), slit viewing, focus monitoring, 
tip-tilt correction and telescope pupil monitoring and correction. In 
addition to this, a temperature stabilized iodine cell is also available 
for providing an accurate velocity reference. The spectrograph will be 
located in a temperature-stabilized box for improved stability.

The light path from M4 to the Coud{\'e} room will be in vacuum tubes -- this
will ensure a minimal level of maintenance of the optics as well as a low 
impact of thermal differences along the light path from the instrument 
room to the dome. The instrument room is expected to be kept at a 
temperature of 15--20\deg C. 

\subsection{Auxiliary instrumentation}

The telescope has two Nasmyth ports of which only one will be used initially.
In order to allow for future instrument upgrades, the telescope will, however,
be designed such that installation of a rotating tertiary mirror will be 
relatively simple. This allows auxiliary instrumentation to be placed at the
second platform. The secondary mirror is designed to allow a field of view of 
10\arcmin$\times$10\arcmin.

%%%%%%%%%%%%%%%%%%%%%%%%%%%%%%%%%%%%%%%%%%%%%%%%%%%%%%%%%%%%%%%%%%%%%%%%%%%
\section{Spectrograph}
%%%%%%%%%%%%%%%%%%%%%%%%%%%%%%%%%%%%%%%%%%%%%%%%%%%%%%%%%%%%%%%%%%%%%%%%%%%

For the asteroseismic observations, the Coud{\'e} spectrograph is the 
principal in\-stru\-ment. In order to determine the required high-precision
velocities an iodine cell will be used for wavelength reference. The 
spectro\-graph will have a (2 pixel) resolution of 120.000 for a 1\arcsec\, 
slit and employ an R4 echelle and a collimated, F/6, beam diameter of 75mm. 
The slit is 10\arcsec\, long, and via the instrument rotator it can 
be rotated in any desired direction allowing the possibility of doing 
asteroseismology for close binary stars such as $\alpha$\,Cen\,A and B.

The spectrograph is designed to cover a wavelength range
of 4800--6800{\AA}. The orders are fully covered at wavelengths shorter
than 5200{\AA}; in the future a larger detector will allow full coverage
of the spectral orders. 

The spectrograph has been designed by P. Span{\`o}, with inspiration from
the design of UVES, HARPS and other modern high-resolution spectrographs. 
It is very compact, roughly $50\times90\times15$cm (without mounts), which
is expected to be an advantage with respect to temperature control. 

In order to provide a well behaved instrumental profile, emphasis was put
on the design of the spectroscopic camera. This resulted in a design 
with an instrumental profile which is nearly diffraction limited over
the entire area of the detector. The detector system will be a 2K$\times$2K
system from Andor of Belfast. This camera has an advanced peltier cooling
system that does not need liquid nitrogen or closed cycle cooling, and 
furthermore the vacuum is ``permanent'' -- so the camera is essentially 
maintenance free. The electronics allow full-frame readout in 5s with 
a readout noise less than  10\,electrons, thus giving only very small overheads
for the spectroscopic time-series observations. The CCD camera is capable
of reading out frames at a speed of 5Mpix. 

\subsection{Spectrograph efficiency and performance}
    
Due to the small wavelength range covered by the spectrograph it is possible
to employ coatings with very high efficiency for both reflective and 
transmissive optics. Many optical companies can deliver optics with
a reflectivity higher than 99.5\% and anti-reflection coatings with 
transmissions better than 99.5\% over the spectrograph design wavelength
interval. Our calculations of the spectrograph efficiency (no slit or 
detector included), show that a throughput in excess of 50\% can be expected. 
With a relatively small telescope and a 75mm beam the slit product is 
a generous 120.000, much higher than for larger telescopes, implying 
smaller slit losses under typical seeing conditions. 

\begin{figure}[!ht]
\plotone{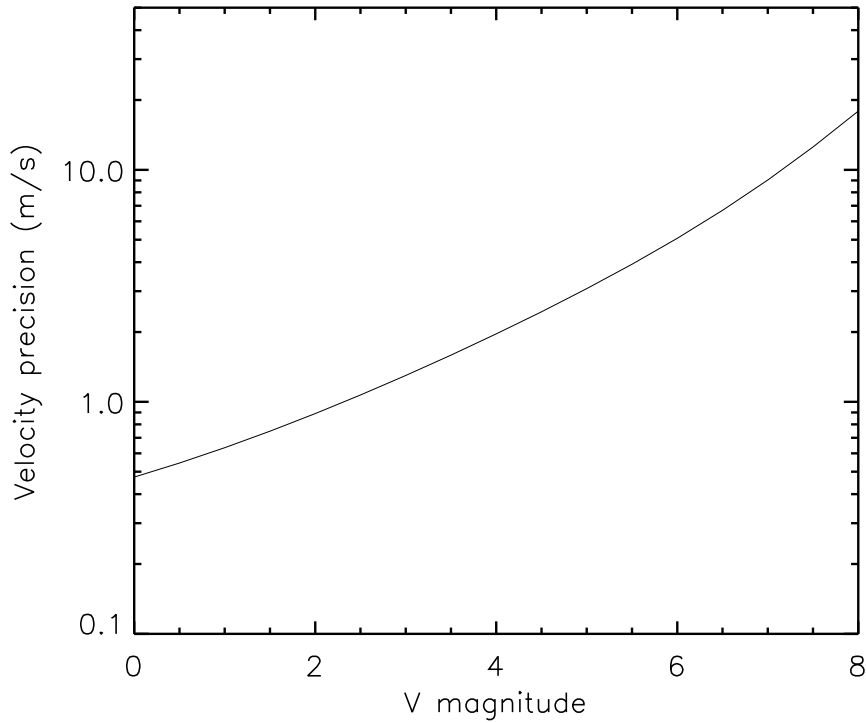}
\caption{
\label{fig:velocity} 
The expected velocity precision for SONG. The calculation is based
on the assumption that the velocity precision scales with signal-to-noise
ratio as found by \cite{iodine2}. We note that at the present time the
{\it i}SONG iodine reduction software has achieved a level of precision of 
$\approx0.8$m/s on UVES archival data for $\alpha$\,Cen\,A; thus with the 
current generation of the software a 1m/s precision can realistically be
expected.
}
\end{figure}

We have carried out calculations of the {\it total} efficiency of the 
spectroscopic system, including all sources of light loss (from atmosphere 
to detector) and these
indicate that in 1\arcsec\, seeing the total system throughput is in 
excess of 8\%.

For the design of the prototype we have included provisions for continuous
control of the location of the telescope pupil on the grating as well as
allowing tip-tilt correction to provide a very stable illumination of the
slit. 

On the assumption that our data reduction code reaches a level of performance
as that of \citet{iodine2} for $\alpha$\,Cen\,A (their Fig.~5) we 
arrive at a predicted velocity precision of better than 1m/s per minute of
observation for stars brighter than $V\approx1$. Stars with lower metallicity, 
shallower lines and higher rotational velocities will not allow such a
high precision to be obtained. In Fig.~\ref{fig:velocity}  the 
calculated velocity precision
vs. stellar magnitude is shown. For stars where a sample time of 1\,minute
(5s read $+$ 55s exposure) is sufficient it will be feasible to carry 
out asteroseismic campaigns to $V\approx6$. 

It should be noted that the iodine cell is not located permanently in the 
light path; therefore ``conventional'' spectroscopy with ThAr calibration
exposures prior to, and after, science exposure will be possible. 
Thus programmes
which do not demand the highest velocity precision are possible in the 
same way as for other telescopes. 

\subsection{Iodine-based velocities with {\it i}SONG}

We have developed an IDL-based data reduction code, called {\it i}SONG, for 
the extraction of
velocities based on iodine-cell observations. At this point, the code has
been tested on the \citet{iodine2} UVES observations of {$\alpha$}\,Cen\,A  
(the raw data were retrieved from the ESO archive) and the SARG observations 
of $\mu$\,Herculis \citep{muher}. For 
the analysis of the {$\alpha$}\,Cen\,A data we achieve a velocity precision of
77cm/s per data point; only slightly poorer than the 70cm/s achieved by 
Butler et al. on the same 688 frames. The current (incomplete) analysis of
the data for $\mu$\,Her indicates that we reach a similar precision as that 
of \citet{muher}. For an example of velocity time-series with
this code we refer the reader to \citet{grundahl1}.

%%%%%%%%%%%%%%%%%%%%%%%%%%%%%%%%%%%%%%%%%%%%%%%%%%%%%%%%%%%%%%%%%%%%%%%%%%%
\section{Summary and status of SONG}
%%%%%%%%%%%%%%%%%%%%%%%%%%%%%%%%%%%%%%%%%%%%%%%%%%%%%%%%%%%%%%%%%%%%%%%%%%%

As of late 2008 the SONG project has obtained full funding for the 
development of a prototype node. Our design revolves around a 
high-performance 1m telescope equipped with a dual-colour lucky-imaging
camera system and a highly efficient high-resolution spectrograph which
will allow 1m/s precision velocities to be obtained for the brightest stars 
in the sky. Better than 10m/s precision is expected for stars brighter than 
$V=6$ per minute of observation. 

SONG is now in its detailed design phase. The optical design is completed
thus allowing final design of the remaining elements. Mechanical design 
of the spectrograph, Nasmyth focal plane and pre-slit systems is
on-going. These systems, with control software, will be tested and integrated
in Aarhus during 2010, with expected delivery to the telescope on the 
first site (Tenerife) in late 2010. The telescope is expected to be 
delivered to the site in mid 2010. 

The ongoing prototype work aside, the main challenge for the coming years
will be to develop the international consortium that will enable the setup
of the full network. Work towards this purpose will be kick-started with
a workshop in Aarhus in March 2009 (see {\tt http://astro.phys.au.dk/SONG}).

%%%%%%%%%%%%%%%%%%%%%%%%%%%%%%%%%%%%%%%%%%%%%%%%%%%%%%%%%%%%%%%%%%%%%%%%%
\acknowledgements 
%%%%%%%%%%%%%%%%%%%%%%%%%%%%%%%%%%%%%%%%%%%%%%%%%%%%%%%%%%%%%%%%%%%%%%%%%

Funding for the SONG project is provided through substantial grants from
the Villum Kann Rasmussen Foundation, The Danish Natural Sciences Research
Council and the Carlsberg foundation. FG wishes to acknowledge many fruitful
conversations with Michael I. Andersen. Henrik K. Bechtold and Anton Norup 
S{\o}rensen are thanked for providing Figs 1 and 2. G. Marcy is thanked 
for advice and encouragement during the development of the {\it i}SONG 
code. S. Leccia and A. Bonanno kindly permitted the use of the raw 
$\mu$\,Her\, data for further tests of the code. 

%%%%%%%%%%%%%%%%%%%%%%%%%%%%%%%%%%%%%%%%%%%%%%%%%%%%%%%%%%%%%%%%%%%%%%%%%
%%  References  %%%%%%%%%%%%%%%%%%%%%%%%%%%%%%%%%%%%%%%%%%%%%%%%%%%%%%%%%
%%%%%%%%%%%%%%%%%%%%%%%%%%%%%%%%%%%%%%%%%%%%%%%%%%%%%%%%%%%%%%%%%%%%%%%%%

%%%%%%%%%%%%%%%%%%%%%%%%%%%%%%%%%%%%%%%%%%%%%%%%%%%%%%%%%%%%%%%%%%%%%%%%%
%%%%%%%%%%%%%%%%%%%%%%%%%%%%%%%%%%%%%%%%%%%%%%%%%%%%%%%%%%%%%%%%%%%%%%%%%
%%%%%%%%%%%%%%%%%%%%%%%%%%%%%%%%%%%%%%%%%%%%%%%%%%%%%%%%%%%%%%%%%%%%%%%%%

\end{document}